\documentclass[12pt]{article}

\usepackage{graphicx}
\usepackage{epstopdf}

\newcommand{\be}{\begin{equation}}
\newcommand{\ee}{\end{equation}}
\newcommand{\ba}{\begin{eqnarray}}
\newcommand{\ea}{\end{eqnarray}}

\newcommand{\beq}{\begin{equation}}
\newcommand{\eeq}{\end{equation}}
\newcommand{\bea}{\begin{eqnarray}}
\newcommand{\eea}{\end{eqnarray}}

\usepackage{graphicx}
 \usepackage{setspace}
 \usepackage{amsfonts}
 \usepackage{amsmath}  
 \usepackage{txfonts} 
\usepackage{color}

\begin{document}
\begin{center}
\vskip 2cm
{\bf \LARGE Black Hole Chemistry}
\vskip 1cm
{\large D. Kubiz\v{n}\'ak\footnote{Email: dkubiznak@perimeterinstitute.ca}
R. B. Mann\footnote{Email: rbmann@uwaterloo.ca}\\
{\it Perimeter Institute for Theoretical Physics\\
35 Caroline St.
Waterloo, Ontario, Canada, N2L 2Y5}\\
\& {\it Department of Physics and Astronomy, University of Waterloo\\
200 University Ave., Waterloo, Ontario N2L 3G1}\\
\bigskip
{\small Essay written for the Gravity Research Foundation 2014 Awards for Essays on Gravitation\\
Submitted March 31, 2014}
}\end{center}

\abstract{The mass of a black hole has traditionally been identified with its energy.  We describe 
a new perspective on black hole thermodynamics, one that identifies the mass of a black hole with chemical  enthalpy,
and   the cosmological constant as thermodynamic pressure. This leads to an understanding of black holes from the viewpoint of chemistry, in terms of concepts such as Van der Waals fluids, reentrant phase transitions, and triple points.   Both charged and rotating black holes exhibit novel  chemical-type phase behaviour, hitherto unseen. }

\newpage 
Our best window as to the  underlying structure of quantum gravity comes from
black hole thermodynamics. By now the parallel with standard thermodynamics is well-known:
\begin{eqnarray}
\textrm{Energy}\quad  E &\leftrightarrow&  M\quad  \textrm{Mass}   \nonumber\\
\textrm{Temperature}\quad  T &\leftrightarrow&  \frac{\kappa}{2\pi} \quad  \textrm{Surface Gravity}   \nonumber\\
\textrm{Entropy}\quad  S &\leftrightarrow&  \frac{A}{4 } \quad  \textrm{Horizon Area}   \nonumber
\label{eq:01}
\end{eqnarray}
in units where $G=c=\hbar =k_B=1$, where on the left-hand side we see the basic thermodynamic quantities of a physical system, and on the right their counterparts in black hole physics.  However this same correspondence is not quite captured by the first law of thermodynamics, which reads 
\begin{eqnarray}
dE= T dS  -PdV + \textrm{work terms}  &\leftrightarrow&  
dM = \frac{\kappa}{8\pi} dA + \Omega dJ + \Phi dQ\,,
 \nonumber
 \label{eq:02}
\end{eqnarray}
where  the $\Omega dJ$ and $\Phi dQ$ terms are understood as thermodynamic work terms.  However there is no counterpart of the ``pressure-volume" term.  Where might it be \cite{Dolan:2012}?

Interesting new developments  have recently taken place that address this question, leading  to a picture in which the mass of a black hole is interpreted as the {\em enthalpy of  spacetime}.  This rather novel idea originates from a consideration of the {\em Smarr relation} \cite{CaldarelliEtal:2000,KastorEtal:2009}.  Originally obtained in $d=4$ 
dimensions \cite{Smarr:1972kt}, it  has the form $(d-3) M = (d-2) TS$ for static asymptotically flat $d$-dimensional
black holes, whose metrics are
\be\label{eq:03}
ds^2 = -f(r) dt^2 + \frac{dr^2}{f(r)} + r^{2} d\Omega_{d-2}^2\,,
\ee
where $f(r)=1 - \frac{1}{d-2}\frac{16\pi}{\omega_{d-2}}\frac{M}{r^{d-3}}+\cdots$, and  $d\Omega_d^2$ is the line element of the unit sphere $S^d$ with a volume $\omega_d =2\pi^{\frac{d+1}{2}}/{\Gamma\bigl(\frac{d+1}{2}\bigr)}$. 
The inclusion of a cosmological constant necessarily modifies the Smarr relation,
as can be shown from a consideration of the scaling properties of the extensive thermodynamic quantities
or from geometric arguments \cite{Dolan:2012,KastorEtal:2009}.  Regarding $M=M(A,\Lambda)$, Euler's theorem
implies that 
\be\label{eq:07}
(d-3) M  = (d-2) \frac{\partial M}{\partial A} A -2 \frac{\partial M}{\partial \Lambda} \Lambda\,,
\ee
and since $T=4\frac{\partial M}{\partial A}$, eq. (\ref{eq:07}) suggests that we regard $P = - \frac{\Lambda}{8 \pi} =\frac{(d-1)(d-2)}{16 \pi \ell^2}$ as a thermodynamic variable \cite{CreightonMann:1995,Padmanabhan:2002sha}, whose conjugate is $V  = -8\pi \frac{\partial M}{\partial \Lambda}$. Putting this together, we have 
\be\label{eq:09}
(d-3) M  = (d-2) TS -2PV  \qquad  {\mbox{and}} \qquad \qquad  dM=TdS+VdP\,,
\ee
as the modified Smarr relation and extended first law of thermodynamics.
We now have the complete thermodynamic correspondence 
\be\label{A1}
\begin{array}{|l|c|l|c|}
\hline
\multicolumn{2}{|c|}{\mbox{Thermodynamics}} & \multicolumn{2}{|c|}{\mbox{Black Hole Physics}} \\
\hline
\mbox{Enthalpy} & H & \mbox{Mass} & M\\
\hline
\mbox{Temperature} & T & \mbox{Surface Gravity} & \frac{  \kappa}{2\pi}\\
\hline
\mbox{Entropy}  &S &\mbox{Horizon Area} & \frac{A}{4} \\
\hline
\mbox{Pressure}  &P & \mbox{Cosmological Constant}  &  -\frac{\Lambda}{8\pi} \\
\hline
\mbox{First Law}  &dH= T dS +VdP + \ldots  & \textrm{First Law} & dM= \frac{\kappa}{8\pi} dA  +VdP + \ldots\\
\hline
\end{array} 
\ee
where the black hole work terms are $\sum_i \Omega_i dJ_i + \Phi dQ$ for multiply rotating and charged black holes. If included then  the more general Smarr relation
\be\label{eq:11}
\frac{d-3}{d-2}M=T S +\sum_i (\Omega^i -\Omega^i_{\infty})J^i -\frac{2}{d-2}PV + \frac{d-3}{d-2}\Phi Q
\ee
holds for all possible $\left[\frac{d}{2}\right]$ rotation parameters, 
where the quantities $\Omega_\infty^i$ allow for the possibility of a rotating frame at infinity  
\cite{Gibbons:2005b, CveticEtal:2010}. 
 
The quantity $P$ is naturally regarded as a thermodynamic pressure, an identification that follows from a cosmological
perspective since a negative cosmological constant induces a vacuum pressure.  The mass $M$ is then understood as a gravitational version of chemical enthalpy, namely the total energy of a system including both its internal energy $E$ and the energy $PV$ required to ``make room for it" by displacing its (vacuum energy) environment: $M=E+PV$. That is, $M$ is the total energy required to ``create a black hole and place it in a cosmological environment''.

The change in perspective on the role of black hole mass and the inclusion of   $\Lambda$ as a pressure term has recently been shown to have a number of remarkable consequences for black hole thermodynamics, making their behaviour analogous to a variety of ``everyday" chemical phenomena.  The first indicator was   the realization that charged black holes behave as Van der Waals fluids \cite{ChamblinEtal:1999a, KubiznakMann:2012}.   Van der Waals equation 
modifies the equation of state for an ideal gas to one that approximates the behaviour of real fluids  
\cite{Goldenfeld:1992}
\be\label{eq:12}
\left(P+\frac{a}{v^2}\right)(v-b)=T  \quad \Leftrightarrow \quad  {P=\frac{T}{v-b}-\frac{a}{v^2}}\,, 
\ee
taking into account the nonzero size of molecules  and the attraction between them, parameterized
by the constants $(b,a)$. Here $v=V/N$ is the specific volume of the fluid, $P$  its pressure, and $T$ its temperature. 
Critical points occur at 
isotherms $T=T_c$, where  $P=P(v)$ has an inflection point  at $P=P_c$ and $v=v_c$, and obey   the universal relation $\frac{P_c v_c}{k T_c}=\frac{3}{8}$ for any such fluid. A liquid/gas phase transition takes place at temperatures
$T<T_c$, and is governed by Maxwell's equal area law, which states that the two phases  coexist when the areas above and below a line of constant pressure drawn through a $P$-$v$ curve are equal.

\begin{figure*}
\centering
\begin{tabular}{cc}
\rotatebox{0}{
\includegraphics[width=0.42\textwidth,height=0.32\textheight]{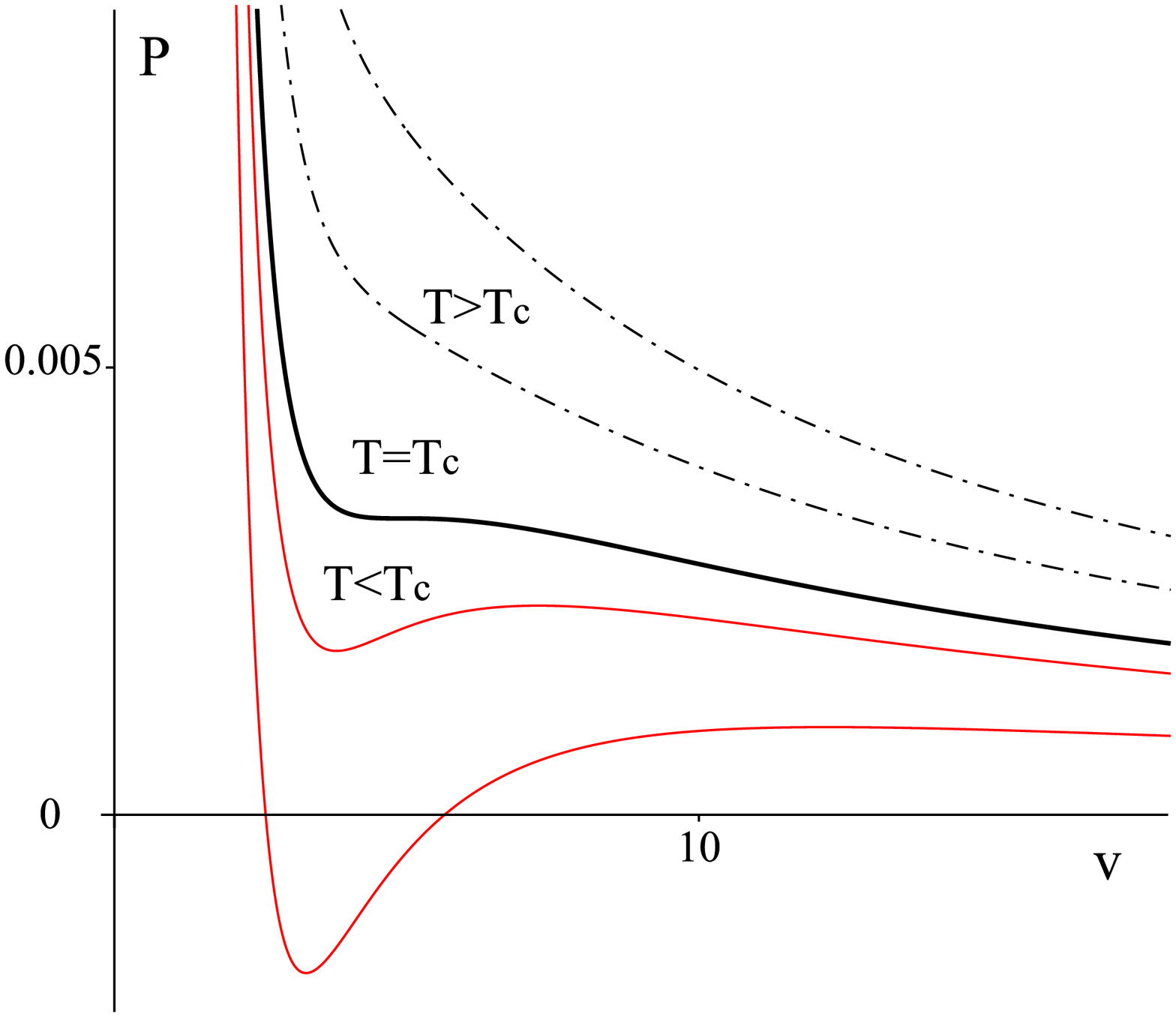}} &
\rotatebox{0}{
\includegraphics[width=0.42\textwidth,height=0.32\textheight]{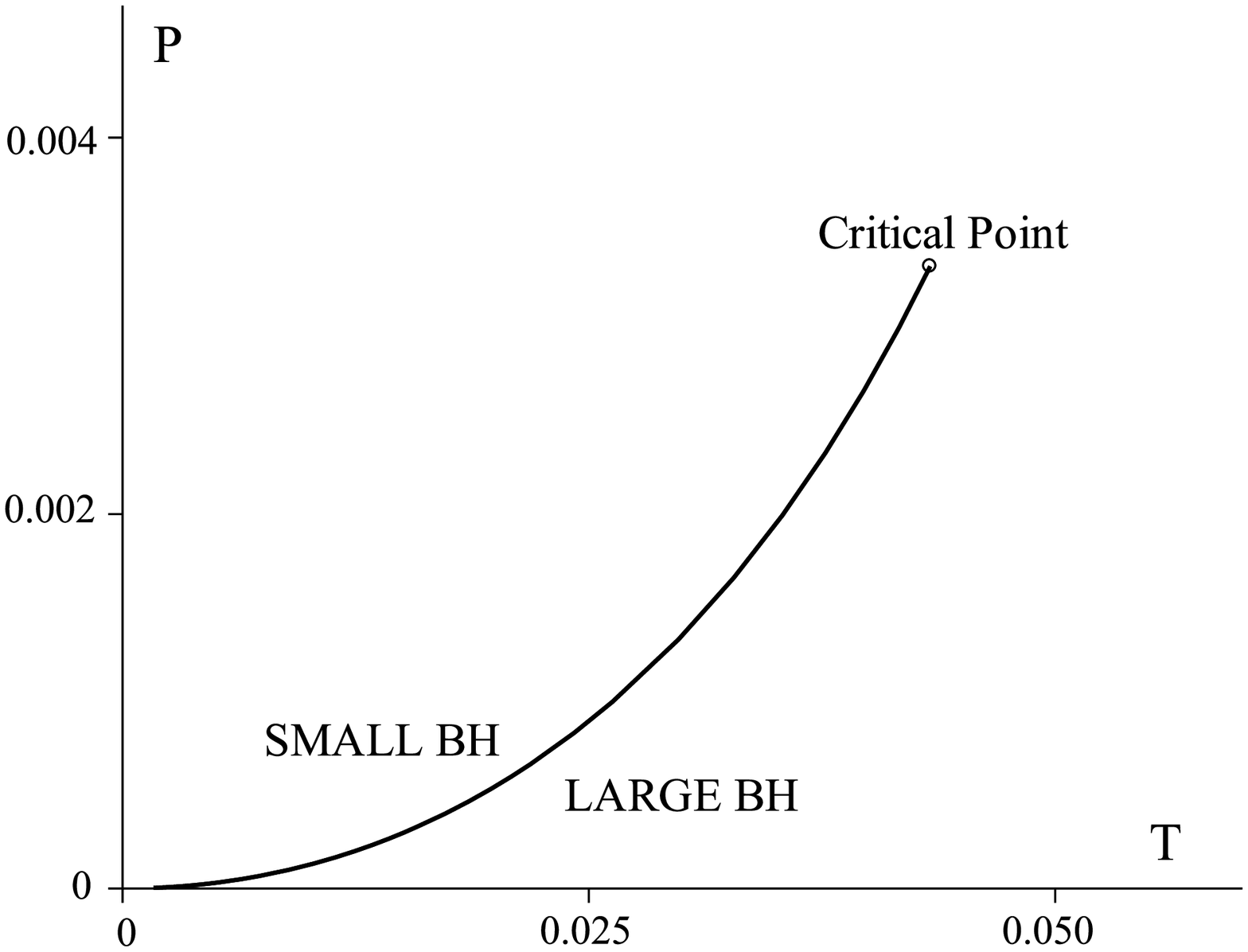}}\\
\end{tabular}
\caption{{\bf 
Fluid behavior of charged AdS black holes.}
{\em Left:} The $P-v$ diagram. Isotherms decrease in temperature from top to bottom. The two upper dark curves correspond to the ``ideal gas'' one-phase behaviour for $T>T_c$, the critical isotherm $T=T_c$ is denoted by the thick dark line, and the lower (red) solid lines correspond to a two-phase state occurring for $T<T_c$ for which the oscillatory part of the isotherm has to be replaced according to the Maxwell's equal area law.
{\em Right:} The $P-T$ diagram. The coexistence line of the two black hole phases terminates at a critical point characterized by the Van der Waals mean field theory critical exponents. In both figures $Q=1$. }
   \label{fig:Pv}
\end{figure*}   
Surprisingly, charged AdS black holes obey the same basic relationships.  
For example, the $d=4$ Reissner-Nordstrom-AdS metric is given by (\ref{eq:03}), with  $f(r) = 1 - \frac{2M}{r} +\frac{Q^2}{r^2}+\frac{r^2}{l^2}$. Its thermodynamic volume is given by the ``Euclidean relationship''
$V  = \frac{4}{3}\pi r_+^3$, where $r_+$ is the location of the event horizon. 
The relation $T = \frac{f'(r_+)}{4 \pi}$ along with the above identification of pressure and volume yield the equation of state
  \be\label{eq:14}
P = \frac{T}{v} - \frac{1}{2\pi v^2} + \frac{2Q^2}{\pi v^4}\,,\qquad v=2r_+ l_P^2=2\left(\frac{3V}{4\pi}\right)^{1/3}\,,
\ee
where  the `specific volume' $v =6V/N$, with $N=A/l_P^2$ counting the number of degrees of freedom associated with the black hole horizon,  $l_P =  \sqrt{G \hbar/c^3}$.  The $P$-$v$ and $P-T$ diagrams, fig.~\ref{fig:Pv},  mimic the behavior of a Van der Waals fluid for any fixed $Q$, with the liquid/gas phase transition replaced by the (first-order) small/large black hole phase transition, still described by the equal-area law.  Interestingly, for the critical point we find  $\frac{P_c v_c}{T_c} = 3/8$, precisely as in the Van der Waals case and the critical exponents  are the same as those for a  Van der Waals fluid \cite{KubiznakMann:2012}; so far no black holes have been found that have different critical exponents.

\begin{figure}
\begin{center}
\rotatebox{0}{
\includegraphics[width=0.42\textwidth,height=0.32\textheight]{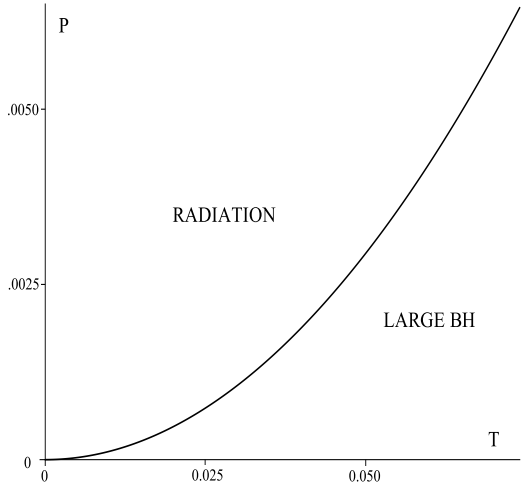}
}
\caption{ {\bf Hawking--Page transition} is a first-order phase transition between thermal radiation in AdS and large stable Schwarzschild-AdS black hole. It occurs when the Gibbs free energy of the black hole becomes lower than that of the thermal radiation. For any pressure $P$ there is a first-order phase transition. 
}
\label{Fig:HPtrans}
\end{center}
\end{figure} 
This approach likewise provides a new perspective on  the renowned 
radiation/large black hole first-order  phase transition first observed by Hawking and Page for Schwarzschild-AdS black holes immersed in a bath of radiation \cite{HawkingPage:1983}. Such a phenomenon  has a dual interpretation for a boundary quantum field theory via the AdS/CFT correspondence and is related to a confinement/deconfinement phase transition in the dual quark gluon plasma.  The coexistence line in the $P-T$ diagram, fig.~\ref{Fig:HPtrans},  has no terminal point, indicating that this phase transition is present for all values  of $\Lambda$, reminiscent of a ``solid/liquid'' phase transition.

\begin{figure*}
\centering
\begin{tabular}{cc}
\rotatebox{0}{
\includegraphics[width=0.42\textwidth,height=0.32\textheight]{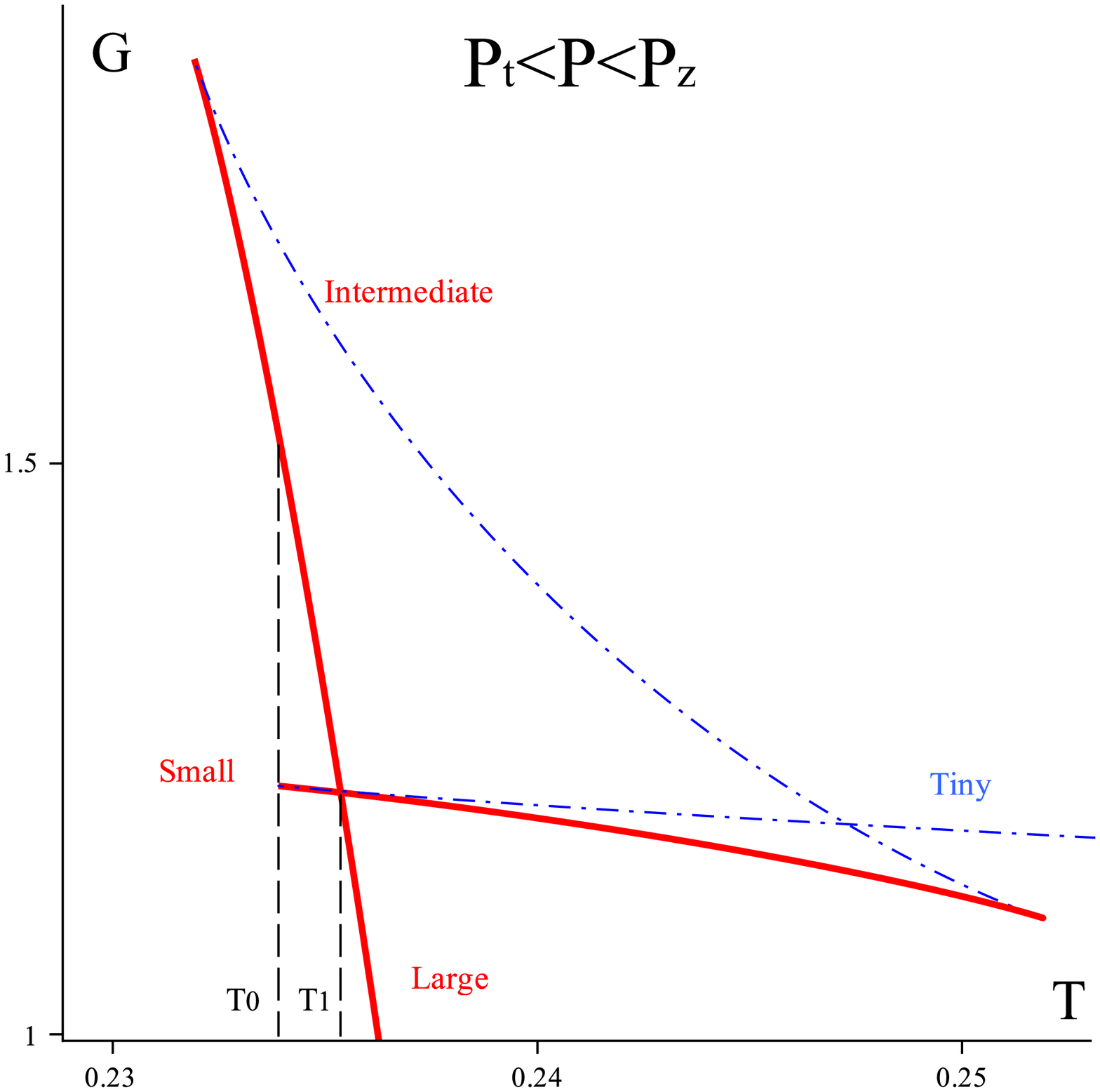}} &
\rotatebox{0}{
\includegraphics[width=0.42\textwidth,height=0.32\textheight]{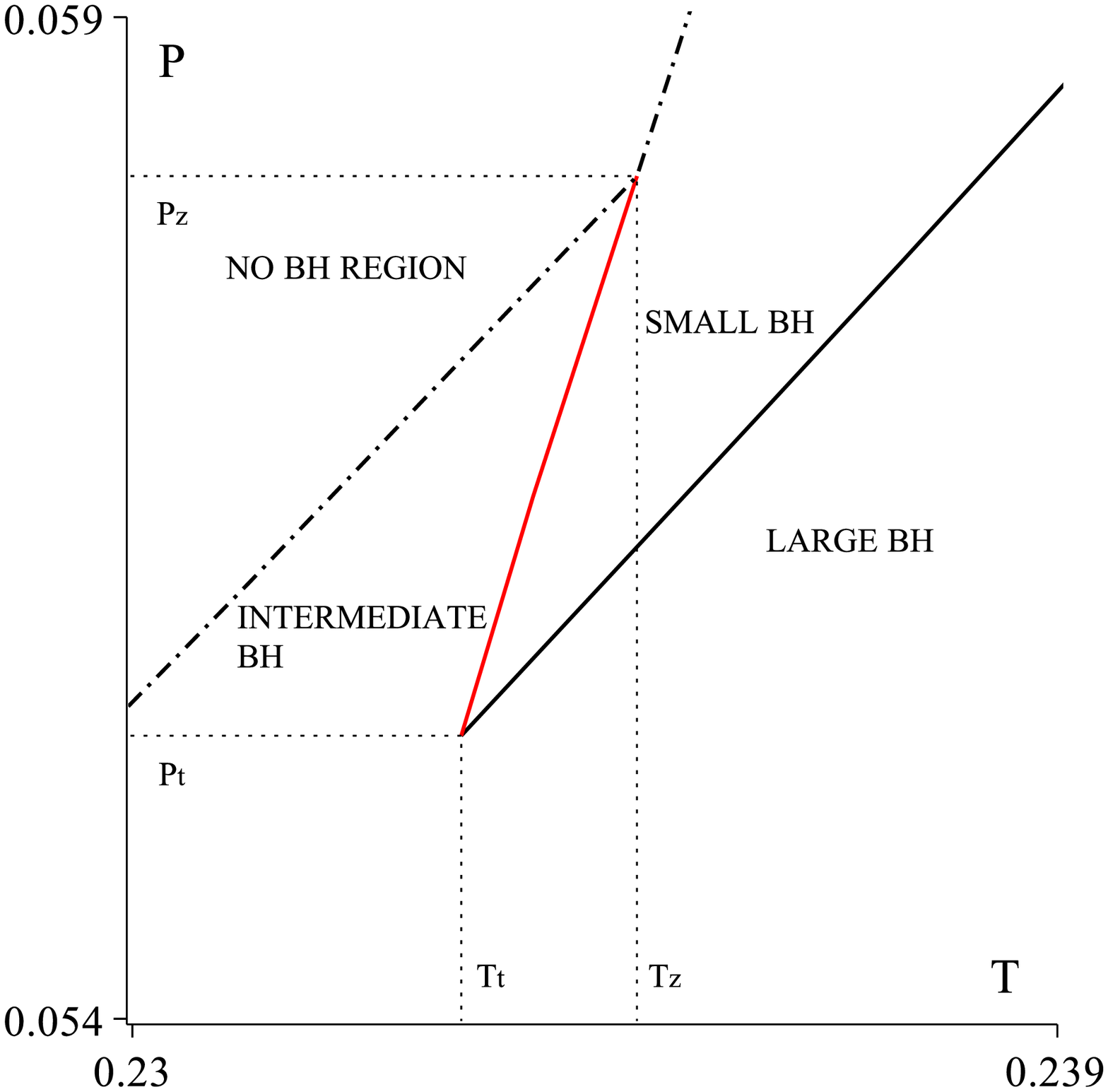}}\\
\end{tabular}
\caption{{\bf Reentrant phase transition.}  {\em Left}:   The Gibbs free energy $G$ of a singly-spinning black hole in $d=6$, displayed for fixed pressure $P\in (P_t,P_z)$ and $J=1$. There is a finite jump in its global minimum indicating the presence of the zeroth-order phase transition. As $T$ decreases, the system follows the lower vertical solid red curve, until at $T=T_1$ it joins the upper horizontal solid red curve corresponding to small stable black holes and undergoes a first order large/small black hole phase transition. As $T$ continues to decrease the system follows this upper curve until $T = T_0$, where G has a discontinuity at its global minimum. Further decreasing $T$, the system jumps to the uppermost vertical red line of large stable black holes. This corresponds to the ``zeroth order"
phase transition between small and large black holes and completes the reentrant phase transition. {\em Right:} The three-phase coexistence diagram displayed in the $P-T$ plane. The first-order phase transition between small/large black holes is displayed by thick solid black curve. It emerges from $(T_t,P_t)$ and terminates at a critical point (not displayed) at $(T_c, P_c)$. The solid red curve indicates the zeroth-order phase transition between intermediate/small black holes; the dashed black curve delineates the `no black hole region'. 
}\label{fig:Gd}
\end{figure*} 
Even more novel thermodynamics emerges for more complicated black hole spacetimes, possibly in higher dimensions \cite{Altamirano:2014tva}. For example a {\em reentrant phase transition} was found for singly-spinning Kerr-AdS black holes in $d\geq 6$ dimensions \cite{Altamirano:2013ane}. 
 First observed in a nicotine/water mixture \cite{Hudson:1904}, this phenomenon  is present in multicomponent fluid systems, gels, ferroelectrics, liquid crystals, and binary gases \cite{NarayananKumar:1994}.  
It refers to a situation in which a system can undergo a  transition from one phase to another and then
back to the first by continuously changing one thermodynamic variable, all others being held constant. 
For $d=6$ black holes the global minimum of the Gibbs free energy suffers from a finite jump (fig.~\ref{fig:Gd}) for
a small range of pressures, $P\in(P_t, P_z)$. Three separate phases of black holes emerge, which we label as intermediate (on the left, a form of large black hole), small  (middle), and large  (on the right).  A standard first order phase transition separates the  small and large black holes,  but  the intermediate and small ones are separated by a finite jump in $G$, indicating a zeroth-order phase transition. For any given pressure in this range, the black hole will change from large to small to large (labeled intermediate) again upon lowering the temperature---giving the reentrant phase transition. Similar behaviour has been observed for Born--Infeld black holes \cite{GunasekaranEtal:2012}.

The gravitational analogue of a {\em triple point} can also be observed once two rotation parameters are included \cite{Altamirano:2013uqa}.
Consider a $d=6$ black hole with rotation parameters  $J_1$ and $J_2$.  We find a variety of phenomena,
dependent on the ratio $q=J_2/J_1$. 
In particular, for $q\ll 1$ a new branch of (locally) stable 
tiny cold black holes appears, with both the $J_2=0$ `no black hole region' and the  unstable branch of tiny hot black holes  disappearing. The zeroth-order phase transition  is `replaced' by a `solid/liquid'-like phase transition of small to large black holes. 
\begin{figure}
\centering
\hspace{1.0cm}
\rotatebox{0}{
\includegraphics[scale=0.5]{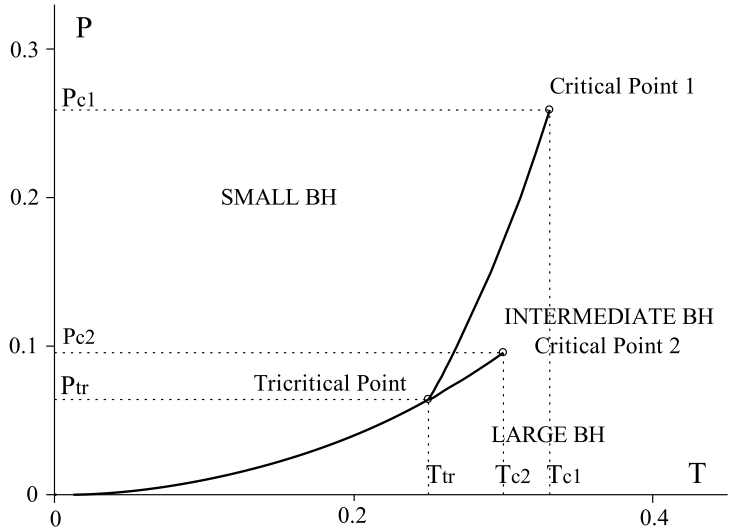}
}
\caption{   {\bf Triple point.} Various black hole phases are displayed for a doubly-spinning Kerr-AdS black hole in $d=6$ and the angular momenta ratio $J_2/J_1=0.05$  in the $P-T$ plane.  The diagram is in many ways analogous to the solid/liquid/gas phase diagram, including the existence of a triple point where three coexistence lines merge together.
Note however that there are two critical points. That is, small/intermediate black hole coexistence line does not extend to infinity (as in the solid/liquid case) but rather terminates, similar to the ``liquid/gas'' coexistence line, in a critical point.
}
\label{fig:6}
\end{figure}

A second critical point and the {\em triple point}  emerge  once $J_2$ becomes sufficiently large. In the range  $q\in (0.00905, 0.0985)$ the phase diagram replicates the behavior of a solid/liquid/gas system, as displayed in fig.~~\ref{fig:6}. Two distinct  first order small/intermediate and intermediate/large black hole phase transitions are possible as the temperature increases for fixed pressure in a certain range---the two transitions terminate at critical points on one end and merge at the other end to form a triple point where the three phases coexist. Contrary to the solid/liquid/gas phase diagram, the small/intermediate black hole coexistence line does not extend all the way to infinity but rather terminates at a critical point. Similar behaviour was subsequently observed for charged AdS black holes in Gauss--Bonnet gravity \cite{Wei:2014hba}.

The physics associated with the extended black hole thermodynamic phase space is only just beginning to be explored.   For example the compressibility $-\frac{1}{V}\frac{\partial V}{\partial P}\vert_{S,J}$ provides
an indicator of the stability of the black hole, with  large compressibility implying a soft equation of state and a system verging on instability \cite{Dolan:2013dga}. $V$ is conjectured to satisfy a reverse isoperimetric inequality \cite{CveticEtal:2010}, a kind of Penrose inequality; if proven valid, this inequality may impose   restrictions on dynamical processes in AdS black hole spacetimes.
 Implications for gauge-gravity duality are also intriguing. Since neither the existence of the reentrant phase transition nor of the triple point depends on a variable $P\sim \Lambda\sim l^{-2}_P N^{-1/2}$, these phenomena will take place for any fixed $P$ within the allowed range. Consequently  reentrant phase transitions and triple points will have  counterparts within the allowed range of $N$ in the dual $SU(N)$ gauge theory.

\medskip
{\small We are grateful to N. Altamirano, B. Dolan, S. Gunasekaran, D. Kastor, Z. Sherkatghanad, 
and J. Traschen for the interesting collaborations that led to the work described here, which was supported in part by the Natural Sciences and Engineering Research Council of Canada and by
the  Perimeter Institute for Theoretical Physics. Research
at Perimeter Institute is supported by the Government of Canada through Industry Canada and by the
Province of Ontario through the Ministry of Research and Innovation.
 }

\bibliographystyle{unsrt}

\end{document}